\documentclass[aps,prl,twocolumn,superscriptaddress]{revtex4-1}
\usepackage{graphicx,hyperref,amsmath,comment,float}
\hypersetup{
     colorlinks   = true,
     allcolors    = [rgb]{0.0,0.0,1.0}
}
\newcommand{\bfpsi}{\mbox{\boldmath$\psi$}}
\newcommand{\bfepsilon}{\mbox{\boldmath$\epsilon$}}
\begin{document}

\title{All-in all-out magnetic order and propagating spin-waves in Sm$_2$Ir$_2$O$_7$}

\author{C. Donnerer}
\affiliation{London Centre for Nanotechnology and Department of Physics and Astronomy, University College London, London WC1E 6BT, United Kingdom}

\author{M. C. Rahn}
\affiliation{Department of Physics, University of Oxford, Clarendon Laboratory, Oxford, OX1 3PU, United Kingdom}

\author{M. Moretti Sala}
\affiliation{ESRF - The European Synchrotron, CS 40220, 38043 Grenoble Cedex 9, France}

\author{J. G. Vale}
\affiliation{London Centre for Nanotechnology and Department of Physics and Astronomy, University College London, London WC1E 6BT, United Kingdom}
\affiliation{Laboratory for Quantum Magnetism, Ecole Polytechnique Federale de Lausanne (EPFL), CH-1015 Lausanne, Switzerland}

\author{D. Pincini}
\affiliation{London Centre for Nanotechnology and Department of Physics and Astronomy, University College London, London WC1E 6BT, United Kingdom}
\affiliation{Diamond Light Source Ltd, Diamond House, Harwell Science and Innovation Campus, Didcot, Oxfordshire OX11 0DE, United Kingdom}

\author{J. Strempfer}
\affiliation{Deutsches Elektronen-Synchrotron DESY, Notkestrasse 85, D-22607 Hamburg, Germany}

\author{M. Krisch}
\affiliation{ESRF - The European Synchrotron, CS 40220, 38043 Grenoble Cedex 9, France}

\author{D. Prabhakaran}
\affiliation{Department of Physics, University of Oxford, Clarendon Laboratory, Oxford, OX1 3PU, United Kingdom}

\author{A. T. Boothroyd}
\affiliation{Department of Physics, University of Oxford, Clarendon Laboratory, Oxford, OX1 3PU, United Kingdom}

\author{D. F. McMorrow}
\affiliation{London Centre for Nanotechnology and Department of Physics and Astronomy, University College London, London WC1E 6BT, United Kingdom}

\begin{abstract}
Using resonant magnetic x-ray scattering we address the unresolved nature of the magnetic groundstate and the low-energy effective Hamiltonian of Sm$_2$Ir$_2$O$_7$, a prototypical pyrochlore iridate with a finite temperature metal-insulator transition. Through a combination of elastic and inelastic measurements, we show that the
magnetic ground state is an all-in all-out (AIAO) antiferromagnet. The magnon dispersion indicates significant electronic correlations and can be well-described by a minimal Hamiltonian that includes Heisenberg exchange ($J=27.3(6)$ meV) and Dzyaloshinskii-Moriya interaction ($D=4.9(3)$ meV), which provides a consistent description of the magnetic order and excitations. In establishing that Sm$_2$Ir$_2$O$_7$ has the requisite inversion symmetry preserving AIAO magnetic groundstate, our results support the notion that pyrochlore iridates may host correlated Weyl semimetals.
\end{abstract}

\maketitle

The search for novel electronic and magnetic phenomena has recently been fruitful in the correlated, strong spin-orbit coupling regime \cite{kim2008novel, kim2009phase, jackeli2009mott, pesin2010mott}. The family of pyrochlore iridates, $R_2$Ir$_2$O$_7$ (where $R$ is a rare-earth element), has received much interest since the prediction of topologically non-trivial states, most prominently the Weyl semimetal (WSM) \cite{yang2010topological, wan2011topological, krempa2012topological, krempa2014correlated}. This is motivated by the observation of metal-insulator transitions as a function of temperature and rare-earth ion radius that occur concomitantly with the onset of magnetic order \cite{yanagishima2001metal, taira2001magnetic, matsuhira2007metal, matsuhira2011metal}. As magnetic order breaks time-reversal symmetry, the WSM state in these correlated materials requires the preservation of inversion symmetry, a scenario distinct from the weakly correlated limit where the opposite is true. Theoretical proposals for the magnetic order with the required symmetries in pyrochlore iridates have focused on the antiferromagnetic all-in all-out (AIAO) structure, where the moments either all point towards or away from the center of the corner shared tetrahedra which form the iridium sublattice. The $R_2$Ir$_2$O$_7$ system thus offers an outstanding opportunity to study novel topological phases in the presence of electronic correlations.

Despite substantial experimental effort, however, the nature of the magnetic order of the Ir ions and the effective spin Hamiltonian have remained elusive in pyrochlore iridates \cite{zhao2011magnetic, tomiyasu2012emergence, shapiro2012structure, disseler2012magnetic2, sagayama2013determination, graf2014magnetism, lefrancois2015anisotropy, clancy2015xray}. Resonant elastic x-ray scattering at the Ir L$_3$ edge of Eu$_2$Ir$_2$O$_7$ has found ${\bf k}={\bf 0}$ magnetic order of undetermined type \cite{sagayama2013determination}. Due to the small magnetic moment of the Ir ion and its high neutron absorption, neutron diffraction has only been successful in studying the rare-earth sublattice. For $R=$ Tb and Nd rare-earths, AIAO magnetic order was found, which has been argued to provide indirect evidence for identical ordering on the Ir lattice \cite{tomiyasu2012emergence, lefrancois2015anisotropy}. An upper limit on the size of the ordered Ir moment was placed at 0.2 $\mu_B$ (Tb) \cite{lefrancois2015anisotropy} and 0.5 $\mu_B$ (Y) \cite{shapiro2012structure}.

\begin{figure}[t]
  \includegraphics[width=\linewidth]{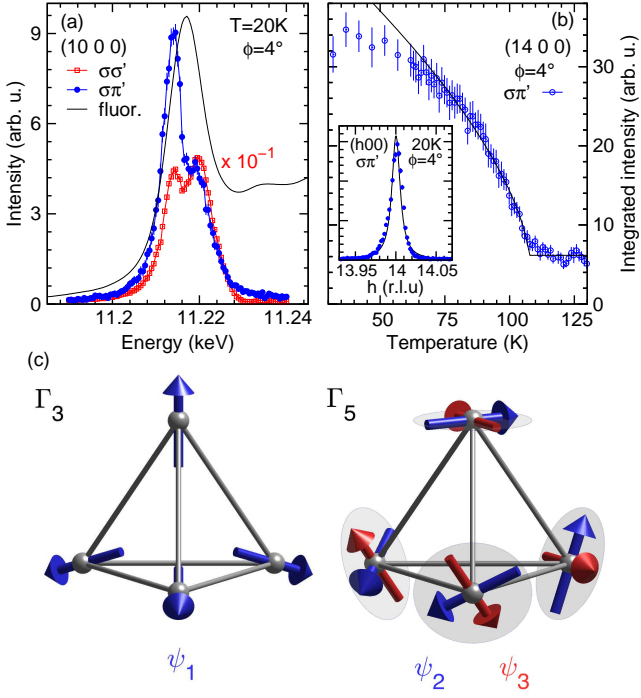}
  \caption{(color online). REXS data showing ${\bf k} = {\bf 0}$ magnetic order in Sm$_2$Ir$_2$O$_7$. (a) Energy dependence of the (10 0 0) reflection in $\sigma\sigma^\prime$ (red squares, ATS resonance) and $\sigma \pi^\prime$ (blue circles, magnetic resonance) polarization channels, plotted with Ir L$_3$ absorption spectrum recorded in total fluorescence mode (black line). The data were corrected for self-absorption. (b) Temperature dependence of the (14 0 0) magnetic reflection in $\sigma \pi^\prime$, revealing $T_N \sim 108$ K. (c) Potential magnetic structures $\Gamma_{3,5}$ proposed by theoretical calculations.}
\label{fig1}
\end{figure}

Here, we use resonant elastic and inelastic x-ray scattering (REXS and RIXS) at the Ir L$_3$ edge to reveal the nature of the magnetic order and excitations of the pyrochlore iridate Sm$_2$Ir$_2$O$_7$. We observe the onset of long-range, ${\bf k}={\bf 0}$ magnetic order at $\sim$ 108 K, close to the metal-insulator transition (MIT) at 114 K. Analysis of the REXS cross-sections constrains the magnetic ground state to be either an AIAO or $XY$ antiferromagnet. Out of these two possibilities, the AIAO structure is the only one consistent with the absence of Goldstone-like modes in the magnon spectrum measured by RIXS. Therefore, the combined REXS and RIXS results show conclusively that the magnetic ground state is AIAO. The magnon dispersion and intensity can be well reproduced by linear spin-wave theory, using a minimal Hamiltonian that includes Heisenberg exchange ($J=27.3(6)$ meV) and Dzyaloshinskii-Moriya interaction ($D=4.9(3)$ meV). The observation of well-defined spinwaves indicates significant electronic correlations in Sm$_2$Ir$_2$O$_7$.

Single crystals of Sm$_2$Ir$_2$O$_7$ were grown by the self-flux method, as described in Ref. \onlinecite{Millican2007crystal}. They were characterised by x-ray diffraction and SQUID magnetometry. In agreement with literature, Sm$_2$Ir$_2$O$_7$ shows a bifurcation in the field-cooled and zero-field-cooled magnetisation at 114 K.

REXS experiments were performed at beamline P09, PETRA III \cite{strempfer2013resonant}. The sample was mounted with the [1 0 0] direction as surface normal. Polarization analysis was performed with a Au $(3 3 3)$ analyser. RIXS experiments were performed at beamline ID20, European Synchrotron Radiation Facility (ESRF) \cite{Moretti2013high}. The overall energy resolution was 25 meV (FWHM). The incident polarization was linear, parallel to the horizontal scattering plane. The incident energy was set to 11.214 keV, 3 eV below the Ir L$_3$ absorption edge maximum.

\begin{figure}
  \includegraphics[width=\linewidth]{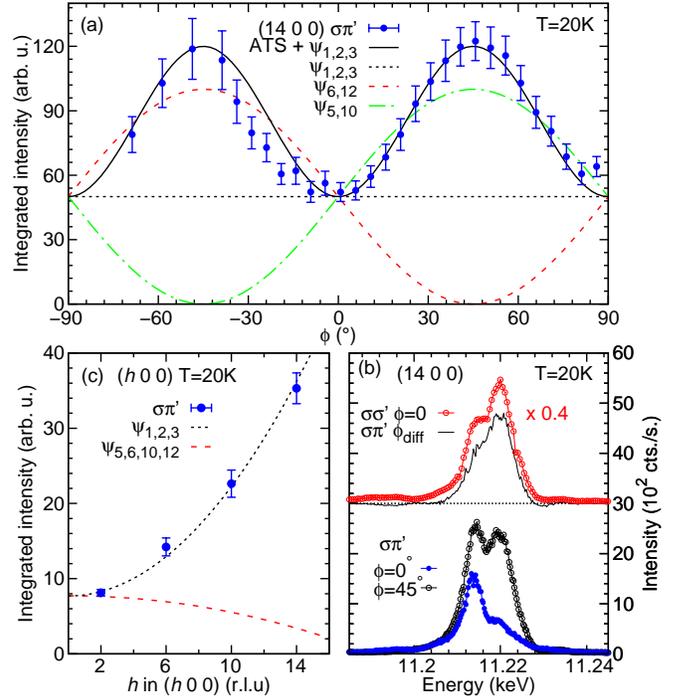}
  \caption{(color online). Analysis of REXS cross-sections of Sm$_2$Ir$_2$O$_7$. (a) Azimuthal dependence of integrated intensity of the (14 0 0) reflection in $\sigma \pi^\prime$. The solid line is the calculated ATS and magnetic scattering $\bfpsi_{1,2,3}$, dotted lines are calculated magnetic contributions of basis vectors $\bfpsi_i$. (b) Energy dependence of the (14 0 0) reflection in $\sigma \pi^\prime$ at azimuthal angles $\phi=0^\circ$ and $\phi=45^\circ$ (bottom). The corresponding difference signal is proportional to the resonance at $\phi=0^\circ$ in the nonmagnetic $\sigma \sigma^\prime$ channel (top). (c) Integrated intensity of charge-forbidden $(h00)$ reflections, plotted with calculated intensities of basis vectors $\bfpsi_i$.}
\label{fig2}
\end{figure}

REXS was previously used to determine ${\bf k} = {\bf 0}$ magnetic order in the pyrochlores Eu$_2$Ir$_2$O$_7$ and Cd$_2$Os$_2$O$_7$ \cite{sagayama2013determination, yamaura2012tetrahedral}. Based on the absence of a structural distortion at the magnetic transition, the AIAO magnetic structure was proposed for both pyrochlores. Here, we extend this approach by exploiting the polarization, momentum and azimuthal dependence of the magnetic scattering to directly characterise the magnetic structure.

Figure \ref{fig1} shows REXS data on magnetic order of Sm$_2$Ir$_2$O$_7$. At 20 K, additional charge-forbidden $(h00)$ reflections (where $h=4n+2$) were discovered. This indicates the formation of ${\bf k} = {\bf 0}$ magnetic order. However, in addition to magnetic scattering, anisotropic tensor of susceptibility (ATS) scattering also contributes to charge-forbidden $(h00)$ reflections \cite{dmitrienko2005polarization}. The ATS contribution in the magnetically-relevant $\sigma \pi^\prime$ channel can be minimized by bringing the $(011)$ direction into the scattering plane (which here defines the azimuthal angle $\phi=0^\circ$) \cite{sagayama2013determination, yamaura2012tetrahedral}. In this scattering geometry, we could identify long-range magnetic order. Fig. \ref{fig1}(a) shows the energy dependence of the (10 0 0) reflection at $\phi=4^\circ$ (corrected for self-absorption), which resonates 3 eV below the absorption maximum, as expected for resonant magnetic scattering in iridates \cite{kim2009phase, boseggia2012antiferromagnetic}. Switching to $\sigma \sigma^\prime$ polarization reveals the shape of the ATS resonance. Two features are seen in the ATS resonance, which were previously attributed to transitions to t$_{2g}$ and e$_g$ levels \cite{sagayama2013determination, ohgushi2013resonant, hirata2013complex}. Fig. \ref{fig1}(b) shows the temperature dependence of the (14 0 0) magnetic reflection together with a power law fit. The data were corrected for beam-heating (see Supplemental Material). The remaining intensity above the ordering temperature ($\sim$ 108 K) can be attributed to weak remnant ATS scattering at $\phi=4^\circ$, as observed in Ref. \onlinecite{ohgushi2013resonant}.

\begin{figure*}[htbp]
  \includegraphics[width=\linewidth]{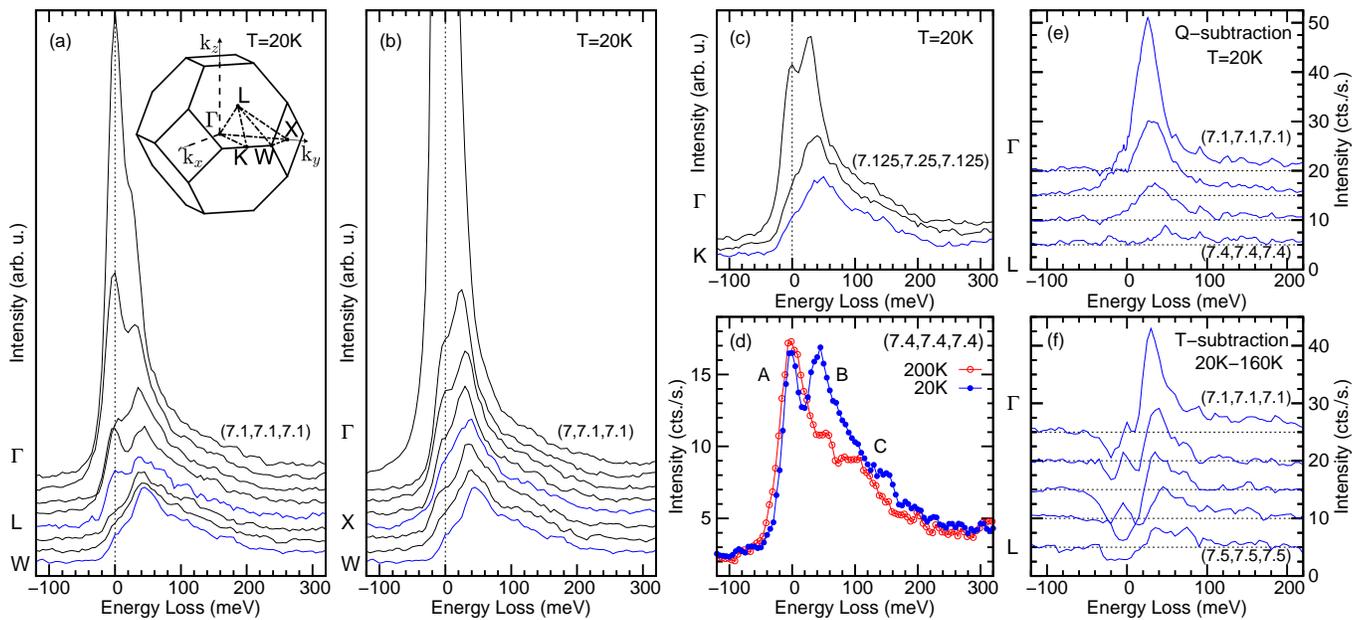}
  \caption{(color online). (a-c) Constant wave vector RIXS energy scans along high-symmetry directions of Sm$_2$Ir$_2$O$_7$ at 20 K. (d) Temperature dependence of magnetic excitations at {\bf Q}={$(7.4,7.4,7.4)$}. (e) Momentum subtraction of RIXS energy scans, the energy scan at the zone boundary $L$ was subtracted from energy scans towards the $\Gamma$ point (elastic line subtracted). (f) Temperature subtraction of RIXS energy scans, the energy scans at 160 K were subtracted from energy scans at 20 K along the $\Gamma-L$ line (elastic line not subtracted).}
\label{fig3}
\end{figure*}

Symmetry-allowed ${\bf k} = {\bf 0}$ magnetic structures can be classified in four irreducible representations $\Gamma_{3,5,7,9}$ with twelve associated basis vectors $\bfpsi_i$ \cite{wills2006magnetic}. The magnetic structure factor can be directly examined by exploiting the REXS cross-sections \cite{hill1996resonant}. Fig. \ref{fig2}(a) shows the integrated intensity of the (14 0 0) reflection in $\sigma \pi^\prime$ polarization (E$_i$=11.214 keV) as a function of azimuthal angle $\phi$. The sinusoidal variation observed originates from ATS scattering ($\propto \sin^2 2 \phi$). Additionally, a temperature dependent intensity offset of similar magnitude to the ATS scattering is present, originating from magnetic scattering. Fig. \ref{fig2}(b) shows that the additional spectral weight at $\phi=45^\circ$ in $\sigma \pi^\prime$ (ATS + magnetic resonance) shares the same energy dependence as the pure ATS resonance in $\sigma \sigma^\prime$, confirming that the magnetic contribution to the resonance remains constant as a function of azimuth. The absence of intensity variation with $\phi$ suggests that the magnetic structure factor lies parallel to the scattering vector. This is also reflected in the intensities of magnetic $(h00)$ reflections which increase with $\sim \sin^2 \theta \propto h^2$ [Fig. \ref{fig2}(c)]. It is only for basis vectors belonging to irreducible representations $\Gamma_3$ ($\bfpsi_1$) and $\Gamma_5$ ($\bfpsi_{2}, \bfpsi_{3}$) that the magnetic structure factor lies parallel to {\bf Q}=$(h00)$. We therefore conclude that the magnetic structure must be either AIAO ($\Gamma_3$, moments along local $z$-axis) or an $XY$ antiferromagnet ($\Gamma_5$, moments within local $xy$-planes). We also note that studying other families of charge-forbidden reflections cannot yield further information to distinguish between these two magnetic structures.

We now consider the magnetic excitations of Sm$_2$Ir$_2$O$_7$ probed by RIXS. Figure \ref{fig3}(a-c) shows constant wave vector energy scans along high-symmetry directions at 20 K. A dispersive feature is observed that reaches a maximum energy of 45 meV at the magnetic ($\equiv$ chemical) Brillouin zone boundaries. Towards the $\Gamma$ point [(7 7 7), charge and magnetic Bragg peak], the intensity of this excitation increases, while the energy decreases to 25 meV. This dispersive feature is temperature dependent, its Lorentizan spectral weight disappears above 90 K at the zone boundaries [Fig. \ref{fig3}(d)]. An additional broad feature however persists at high temperature, extending from a quasi-elastic response to $\sim$ 200 meV. This broad feature appears independent of momentum transfer and temperature: Subtracting the energy scan at the zone boundary ($L$) from energy scans towards the zone center ($\Gamma \rightarrow L$) at 20 K reveals that only the inelastic Lorentzian feature disperses [Fig. \ref{fig3}(e)]. Similarly, subtracting energy scans at 160 K from energy scans at 20 K reveals that the additional spectral weight at low temperature consists predominantly of a low-energy Lorentzian component [Fig. \ref{fig3}(f)]. This suggests that the low-energy RIXS response of Sm$_2$Ir$_2$O$_7$ comprises three features: An elastic signal (at 0 meV, feature A), a gapped, dispersive, temperature-dependent Lorentzian feature (at 25-45 meV, feature B) and a non-dispersive, temperature-independent continuum of excitations (at 0-200 meV, feature C).

The temperature dependence of feature B suggests a magnetic origin. The magnetic excitations of AIAO order in pyrochlore iridates are predicted to be gapped ($\sim$100 meV) with a dispersion bandwidth of 15 meV \cite{lee2013magnetic}. While the spin gap we observe is smaller ($\sim$25 meV), the dispersion bandwidth is roughly consistent ($\sim$20 meV). The origin of feature C is unclear. A similar feature has been observed in the metallic Pr$_2$Ir$_2$O$_7$ and was interpreted as paramagnetic fluctuations \cite{clancy2015xray}. An alternative interpretation of feature C are lattice vibrations. Raman scattering of Sm$_2$Ir$_2$O$_7$ \cite{hasegawa2010raman} has revealed multiple single-phonon modes that, when including higher harmonics, would appear as a broad continuum of excitations within the RIXS energy resolution. This scenario was also proposed to occur in Na$_2$IrO$_3$ \cite{gretarsson2013magnetic}. Based on these considerations, the low-energy RIXS response of Sm$_2$Ir$_2$O$_7$ was fitted with a delta function at zero energy loss for elastic scattering (feature A), a Lorentzian function for the single magnon excitation (feature B) and an antisymmetrized Lorentzian for feature C (see Supplemental Material). All functions were convoluted with the instrumental resolution. The functional form of feature C was chosen on phenomenological grounds, its width and center were determined by a global refinement and then fixed for individual {\bf Q}-points.

\begin{figure}
  \includegraphics[width=\linewidth]{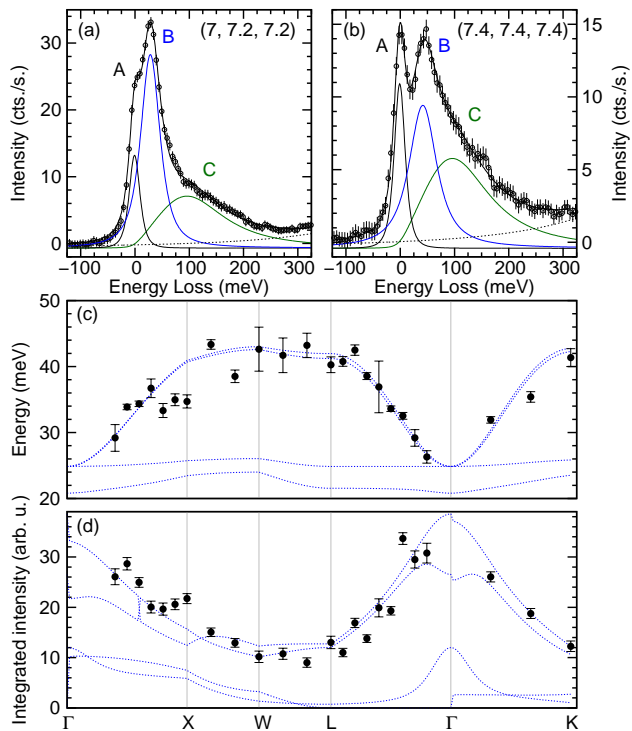}
\caption{(color online). (a-b) Representative fitted RIXS energy scans of Sm$_2$Ir$_2$O$_7$. (c) Fitted energy dispersion of feature B as a function of momentum transfer (black dots) along with calculated magnon dispersion of AIAO order (blue dotted lines). (d) Fitted intensity dispersion of feature B (black dots) along with calculated total dynamical structure factor $S({\bf q},\omega)=\sum_a S^{aa}({\bf q},\omega)$ of AIAO order (blue dotted lines).}
\label{fig4}
\end{figure}

To describe the magnetic excitations, we employ the minimal Hamiltonian $H = \sum_{ij} ( J {\bf S_i} \cdot {\bf S_j} + {\bf D}_{ij} \cdot {\bf S_i} \times {\bf S_j} )$. Here we have included Heisenberg exchange and the Dzyaloshinskii-Moriya interaction (DMI). Previous literature has also included an anisotropic exchange term \cite{pesin2010mott, krempa2012topological, lee2013magnetic}. As this term is expected to be small, we exclude it to reduce the number of free parameters. The sign of the DMI determines which ground state is realized \cite{elhajal2005ordering}: A \emph{direct} (positive) DMI will favor the AIAO ground state, whereas the \emph{indirect} (negative) DMI stabilises the $XY$ ground state (amongst others). We used SpinW \cite{toth2015linear} to calculate the spin-wave spectrum in the Holstein-Primakoff approximation.

Figure \ref{fig4} shows the fitted dispersion of magnetic excitations. For the $XY$ pyrochlore antiferromagnet and $D < 0$, pseudo-Goldstone modes are expected \cite{ruff2008spin, savary2012order, zhitomirsky2012quantum}. This can be understood from the continuous, classical degeneracy of groundstates, characterized by the arbitrary angle of the moments within the local $xy$-planes. In our data, the magnetic excitations are however gapped throughout the Brillouin zone. While we are limited by the experimental resolution, close to the $\Gamma$ point a gapped mode that carries the majority of spectral weight is observed [Fig. \ref{fig4}(a)], indicating that Goldstone-like modes are absent. An $XY$ antiferromagnetic ordering therefore appears incompatible with the excitations we observe.

Conversely, for AIAO order and $D > 0$, all four magnetic modes are gapped  \cite{lee2013magnetic}. In our data, only one dispersive mode can be distinguished. This mode was fitted with $J=27.3(6)$ meV and $D=|D_{ij}|=4.9(3)$ meV. For these parameters, two quasi-degenerate modes disperse from 25 meV to 45 meV, carrying the majority of the spectral weight. While two additional modes are predicted in the calculation, their absence in the data can be explained by their small spectral weight and low energy (20-25 meV) that will make them indistinguishable from an elastic signal.

Linear spin-wave theory implicitly assumes a strong-coupling limit. In Ref. \onlinecite{lee2013magnetic}, the magnetic excitations have been explicitly calculated as a function of the Hubbard interaction $U$. This allows direct comparison with our data: We observe well-defined spin-waves across the entire Brillouin zone that disperse only towards the $\Gamma$ point, which is only compatible with large Hubbard $U$ calculations. However, the fits to the data reveals an intrinsic Lorentzian FWHM of 22(2) meV near the $\Gamma$ point, on the order of the dispersion bandwidth. Indeed, Ref. \onlinecite{lee2013magnetic} predicts that the magnetic excitations become strongly damped as $U$ is decreased. This indicates intermediate-to-strong correlations in Sm$_2$Ir$_2$O$_7$.

It is also interesting to consider the fitted exchange constants. For $J_{\text{eff}} = 1/2$ moments, larger values for the DMI were predicted ($D/J \approx 0.63$ in Ref. \onlinecite{pesin2010mott}, whereas we find $D/J \approx 0.18$). Crystal-field excitations show that trigonal distortions are as strong as the spin-orbit coupling in pyrochlore iridates (both $\sim$ 400 meV, data not shown, see Refs. \onlinecite{hozoi2014longer, uematsu2015large}). This places the ground state halfway between a $J_{\text{eff}} = 1/2$ and a $S=1/2$ doublet and may explain the smaller exchange anisotropy. 

In summary, our study has found compelling evidence for AIAO magnetic order on the Ir sublattice of Sm$_2$Ir$_2$O$_7$. While this has been identified as a prerequisite for stabilizing Weyl semimetals, the magnetic excitation spectrum indicates significant electronic correlations. This may place Sm$_2$Ir$_2$O$_7$ in a topologically trivial Mott limit. ARPES has recently shown that in the paramagnetic, metallic Pr$_2$Ir$_2$O$_7$ quadratic band touching occurs at the Fermi level \cite{kondo2015quadratic}. It will therefore be of great interest to explore the physics in between these limits by studying magnetically-ordered pyrochlore iridates closer to the quantum critical point \cite{savary2014new}.\\

We would like to thank S. Toth, L. Hozoi, S. Nishimoto, R. Yadav, M. Pereiro and J. van den Brink for fruitful discussions. This work is supported by the UK Engineering and Physical Sciences Research Council, under Grants No. EP/J016713/1 and No. EP/J017124/1. Parts of this research were carried out at the light source PETRA III at DESY, a member of the Helmholtz Association (HGF).

\section{Supplemental Material}

We provide supplementary material on 1) the crystal structure, 2) the resonant magnetic x-ray cross-sections, 3) anisotropic tensor of susceptibility (ATS) cross-sections, 4) combined resonant elastic x-ray scattering (REXS) cross-sections, 5) REXS beam heating correction, 6) fitting of resonant inelastic x-ray scattering (RIXS) data and 7) linear spin-wave theory calculations.

\section{S1. Crystal structure of S\lowercase{m}$_2$I\lowercase{r}$_2$O$_7$}

Sm$_2$Ir$_2$O$_7$ crystallizes in the cubic pyrochlore structure, spacegroup $Fd\bar{3}m$ (No. 227, origin choice 2), with lattice constant $a=10.3$ \r{A} at room temperature. 

\section{S2. Resonant magnetic x-ray scattering}

Symmetry-allowed ${\bf k} = {\bf 0}$ magnetic structures of Ir at $16c$ can be classified in four irreducible representations $\Gamma_{3,5,7,9}$ with twelve associated basis vectors $\psi_i$ \cite{wills2006magnetic}. The magnetic structure factor at wavevector ${\bf Q}$ can be written as:

\begin{equation}
{\bf M} ({\bf Q}) = \sum_{j=1}^4 {\bf m}_j e^{ 2 i \pi {\bf Q} \cdot {\bf r}_j }
\end{equation}

where the sum $j$ runs over the four inequivalent Ir sites and ${\bf m}_j$ the magnetic Fourier components. For charge-forbidden $(h00)$ reflections, the magnetic structure factor simplifies to:

\begin{equation}
{\bf M} {(h00)} = {\bf m}_1+{\bf m}_2-{\bf m}_3-{\bf m}_4 
\end{equation}

Table \ref{SFs} shows the calculated magnetic structure factors using the basis vectors defined in Ref. \onlinecite{wills2006magnetic}. The three $S$-domains for $\Gamma_5$ are included, which arise from the degeneracy of the crystallographic axes.

\begin{table}[tbh]
\begin{tabular} { c c c l }
  IR  &  domain & BV  &   ${\bf M} {(h00)}$ \\
\hline
$\Gamma_{3}$ & 1   & $\bfpsi_{1}$ & $(\frac{4}{\sqrt{3}}, 0, 0)$ \\
$\Gamma_{5}$ & 1   & $\bfpsi_{2}$ & $(\frac{8}{\sqrt{6}}, 0, 0)$ \\
			&      & $\bfpsi_{3}$ & $(0, 0, 0)$ \\
		       	& 2   & $\bfpsi_{2}$ & $(\frac{-4}{\sqrt{6}}, 0, 0)$ \\
			&      & $\bfpsi_{3}$ & $(\frac{-4}{\sqrt{2}}, 0, 0)$ \\
		       	& 3   & $\bfpsi_{2}$ & $(\frac{-4}{\sqrt{6}}, 0, 0)$ \\
			&      & $\bfpsi_{3}$ & $(\frac{4}{\sqrt{2}}, 0, 0)$ \\
$\Gamma_{7}$ & 1   & $\bfpsi_{4}$ & $(0, 0, 0)$ \\
			&      & $\bfpsi_{5}$ & $(0, 0, \frac{4}{\sqrt{2}})$ \\
		        &      & $\bfpsi_{6}$ & $(0,\frac{-4}{\sqrt{2}}, 0)$ \\
$\Gamma_{9}$ & 1   & $\bfpsi_{7}$ & $(0, 0, 0)$ \\
			&      & $\bfpsi_{8}$ & $(0, 0, 0)$ \\
		        &      & $\bfpsi_{9}$ & $(0,0, 0)$ \\ 
		        &      & $\bfpsi_{10}$ & $(0, 0, \frac{4}{\sqrt{2}})$ \\
		        &      & $\bfpsi_{11}$ & $(0,0,0)$ \\ 
		        &      & $\bfpsi_{12}$ & $(0,\frac{4}{\sqrt{2}},0)$ \\ 
\end{tabular}
\caption{Magnetic structure factors for basis vectors $\bfpsi_{i}$ (defined in Ref. \cite{wills2006magnetic}) for charge-forbidden $(h00)$ reflections (where $h=4n+2$).}
 \label{SFs}
\end{table}

The amplitude for resonant magnetic scattering at the Ir L$_3$ edge (dipole transitions) can be written as:

\begin{equation}
A_{m} = -i F_{m} ( {\bfepsilon^\prime} \times {\bfepsilon }) \cdot {\bf M} ({\bf Q}) 
\end{equation}

where ${\bfepsilon^{(\prime)}}$ are the incoming (outgoing) polarization and $F_{m}$ describes the strength of the magnetic resonance. This can be cast in the $2 \times 2$ Jones matrix \cite{hill1996resonant}:

\begin{equation}
A_{m} = -i F_{m}
\begin{pmatrix}
0 & M_1^\prime \cos \theta + M_3^\prime \sin \theta \\
M_3^\prime \sin \theta - M_1^\prime \cos \theta & - M_2^\prime \sin 2 \theta \\
\end{pmatrix}
\end{equation}

where $M_i^\prime$ are the components of ${\bf M} ({\bf Q}) $ in the laboratory frame (as defined in Ref. \onlinecite{hill1996resonant}) and $\theta$ the Bragg angle. In $(h00)$ scattering condition, the crystallographic axes map onto the laboratory frame as:

\begin{equation}
{\bf M}
= {\frac{1}{\sqrt{2}}}
\begin{pmatrix}
0 & 0 & -\sqrt{2} \\
\sin \phi +  \cos \phi & \cos \phi - \sin \phi& 0 \\
\cos \phi - \sin \phi & -\cos \phi - \sin \phi & 0 \\
\end{pmatrix}
{\bf M^\prime}
\end{equation}

where the azimuthal angle $\phi$ is defined $\phi = 0^\circ$ when the (011) direction is parallel to the scattering plane. Hence we can write the amplitude for resonant magnetic scattering in the rotated polarisation ${\sigma \pi^\prime}$ channel as:

\begin{table}[tbh]
\begin{tabular} {c c c c }
  IR  &  domain & BV  &   $A_m^{\sigma \pi^\prime} {(h00)}$ \\
\hline
$\Gamma_{3}$ & 1  & $\bfpsi_{1}$ & $i F_{m} \frac{4}{\sqrt{3}} \sin \theta $ \\
$\Gamma_{5}$ & 1  & $\bfpsi_{2}$ & $i F_{m} \frac{8}{\sqrt{6}} \sin \theta $ \\
			&      & $\bfpsi_{3}$ & $0$ \\
		       	& 2  & $\bfpsi_{2}$  & $-i F_{m} \frac{4}{\sqrt{6}} \sin \theta $ \\
			&      & $\bfpsi_{3}$ & $-i F_{m} \frac{4}{\sqrt{2}} \sin \theta $ \\
		       	& 3  & $\bfpsi_{2}$  & $-i F_{m} \frac{4}{\sqrt{6}} \sin \theta $ \\
			&      & $\bfpsi_{3}$ & $ i F_{m} \frac{4}{\sqrt{2}} \sin \theta $ \\
$\Gamma_{7}$ & 1  & $\bfpsi_{4}$  & $0$\\
			&      & $\bfpsi_{5}$ & $-2i F_{m} (\cos \phi - \sin \phi)\cos \theta $ \\
			&      & $\bfpsi_{6}$ & $2i F_{m}  (\cos \phi + \sin \phi)\cos \theta $ \\
$\Gamma_{9}$ & 1  & $\bfpsi_{7}$  & $0$\\
			&      & $\bfpsi_{8}$ & $0$ \\
			&      & $\bfpsi_{9}$ & $0$ \\
			&      & $\bfpsi_{10}$ & $-2i F_{m} (\cos \phi - \sin \phi)\cos \theta $ \\
			&      & $\bfpsi_{11}$ & $0$ \\		
			&      & $\bfpsi_{12}$ & $-2i F_{m} (\cos \phi + \sin \phi)\cos \theta $ \\
\end{tabular}
\caption{Resonant magnetic scattering amplitudes for charge-forbidden $(h00)$ reflections, where $h=4n+2$, for basis vectors $\bfpsi_{i}$ (defined in Ref. \cite{wills2006magnetic}).}
\end{table}

\section{S3. ATS scattering}

The contribution of ATS scattering on charge-forbidden $(h00)$ reflections can be written as \cite{dmitrienko2005polarization}:

\begin{equation}
A_{ATS}^{(h00)} = F_{ats}
\begin{pmatrix}
\cos 2 \phi & - \sin \theta \sin 2 \phi \\
\sin \theta \sin 2 \phi & \sin^2 \theta \cos 2 \phi \\
\end{pmatrix}
\end{equation}

where $F_{ats}$ describes the strength of the ATS resonance.

\section{S4. REXS cross sections}

We can now construct the REXS cross section involving magnetic and ATS contributions for charge-forbidden $(h00)$ reflections. For all-in all-out magnetic order ($\bfpsi_1$) we find:

\begin{equation}
\resizebox{0.95\hsize}{!}{$
A^{(h00)}_{REXS} = \\
\begin{pmatrix}
F_{ats} \cos 2 \phi & \sin \theta (i \frac{4}{\sqrt{3}} F_m  -  F_{ats} \sin 2 \phi)  \\
\sin \theta (F_{ats} \sin 2 \phi +i \frac{4}{\sqrt{3}} F_{m} ) & F_{ats} \sin^2 \theta \cos 2 \phi \\
\end{pmatrix}$}
\end{equation}

Then the intensity in the $\sigma \pi^\prime$ polarization channel is:

\begin{equation}
I_{\sigma \pi}^{h00} (\phi) = \sin^2 \theta \left(\frac{16}{3} F_{m}^2 + \sin^2 \phi F_{ats}^2 \right)
\end{equation}

which at $\phi=0^\circ$ simplifies to

\begin{equation}
I_{\sigma \pi}^{h00} (\phi=0^\circ) = \frac{16}{3} F_{m}^2 \sin^2 \theta \propto h^2
\end{equation}

at constant wavelength. 

\section{S5. REXS beam heating correction}

\begin{figure}[tbh]
  \includegraphics[width=\linewidth]{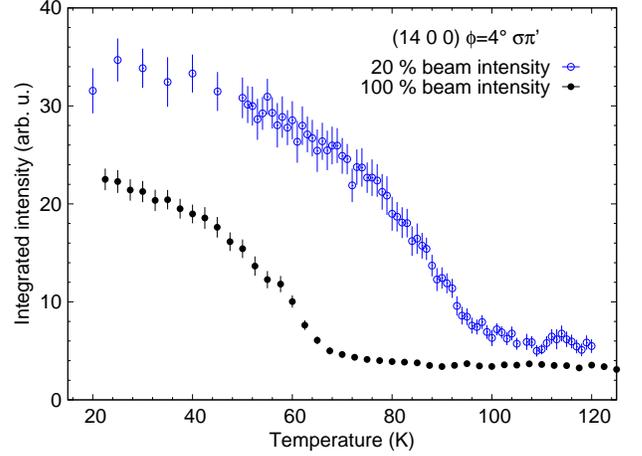}
  \caption{(color online) REXS temperature dependence of magnetic (14 0 0) reflection showing substantial beam heating. The black, filled data points are taken without attenuation, the blue, open data points are taken with 20\% beam intensity.}
\label{fig1s}
\end{figure}

A strong beam heating effect was found in the temperature dependence of the resonant magnetic x-ray scattering. Figure \ref{fig1s} shows the temperature dependence of the magnetic (14 0 0) reflection with full beam intensity and attenuated to 20\% beam intensity. Heating in excess of 30 K at nominal 90 K is seen. In order to quantify any remaining beam heating at 20\% beam intensity we further attenuated the beam to 1\% of the full intensity at a nominal temperature of 90 K. With 1\% beam intensity, the integrated intensity of a Lorentzian fit to a rocking curve at 90 K increased by a factor of $\sim$ 1.57 compared to 20\% beam intensity. As it is not feasible to perform a temperature dependence with 1\% beam intensity, we used the fitted power law to estimate a beam heating of $\sim$ 11.8 K at 20\% beam intensity at nominal 90 K and corrected the data accordingly. Precise estimates of the critical exponent and transition temperature can therefore not be made with this data.

\section{S6. Fitting of RIXS data}

Constant wave vector energy scans were fitted with a delta function (elastic scattering, feature A), a Lorentzian (single magnon excitation, feature B) and an antisymmetrized Lorentzian (feature C), where the latter is defined as:

\begin{equation}
L (E) = \frac{\Gamma}{(E-E_0) + \Gamma^2} -  \frac{\Gamma}{(E+E_0) + \Gamma^2}
\end{equation}

where $E_0$ is the position of the maximum and $\Gamma$ the width. All features were convoluted with the instrumental resolution function that can be well described by a Pearson VII function with FWHM = 23.4(2) meV and shape parameter $\mu=2.0(1)$ (where $\mu=1$ corresponds to a Lorentzian and $\mu=\infty$ to a Gaussian).

The center and width of feature C were determined by a global refinement of all {\bf Q}-points $i$ that minimizes $\sum_{i=1}^{n} \chi^2_i$. These parameters were subsequently fixed for determining the dispersion of feature B. This minimizes the number of fitted parameters and ensures that dispersing spectral weight is described by feature B.

\section{S7. Linear spin-wave theory}

We used SpinW \cite{toth2015linear} to calculate the magnon dispersion in the Holstein Primakoff approximation. We used the following Hamiltonian:

\begin{equation}
H = \sum_{ij} ( J {\bf S_i} \cdot {\bf S_j} + {\bf D}_{ij} \cdot {\bf S_i} \times {\bf S_j} )
\end{equation}

\begin{figure}[H]
  \includegraphics[width=\linewidth]{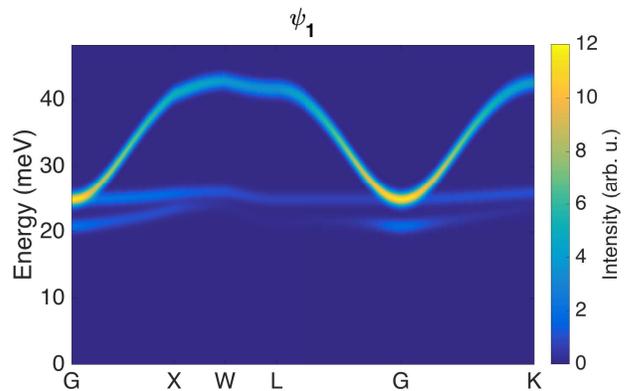}
  \caption{Calculated magnon dispersion of the AIAO antiferromagnet ($\bfpsi_1$) using $J=27.3$ meV and $D=4.9$ meV, within the Brillouin zone G = (7 7 7).}
\label{fig2s}
\end{figure}

\begin{figure}[H]
  \includegraphics[width=\linewidth]{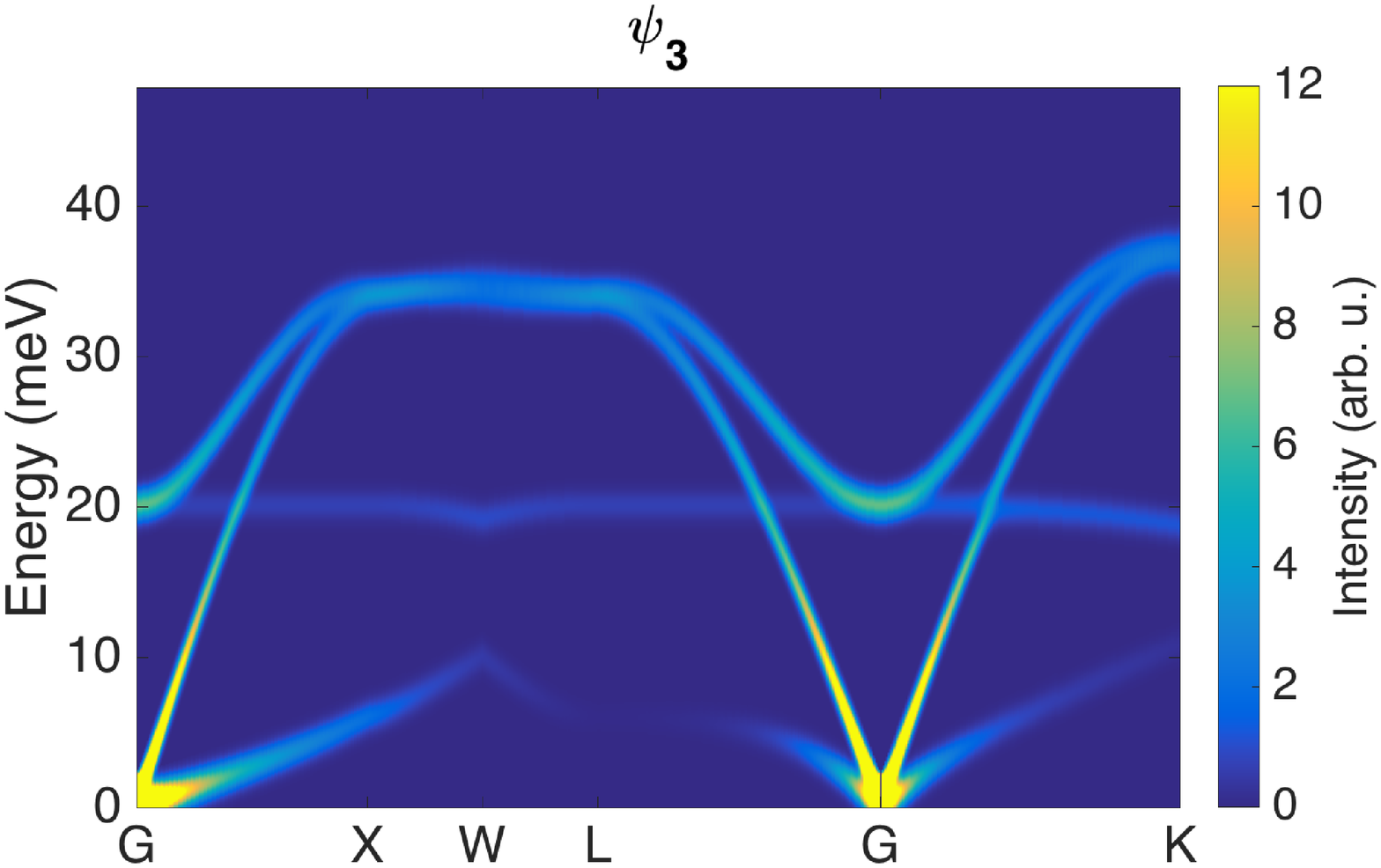}
  \caption{Calculated magnon dispersion of an $XY$ antiferromagnet ($\bfpsi_3$) using $J=27.3$ meV and $D=-4.9$ meV, within the Brillouin zone G = (7 7 7).}
\label{fig3s}
\end{figure}

Here we have included Heisenberg exchange and the Dzyaloshinskii-Moriya interaction (DMI). The all-in all-out (AIAO, $\bfpsi_1$) and $XY$ ($\bfpsi_{2,3}$) antiferromagnetic structure could be stabilized by using positive and negative value of $D$, respectively. Figures \ref{fig2s} and \ref{fig3s} show the calculated magnon dispersions of the magnetic structures $\bfpsi_1$ and $\bfpsi_3$ using $J=27.3$ meV and $D=\pm4.9$ meV.

\end{document}